%% file: article.tex
\documentclass[
  reprint,
  aps,
  pra,
  amsmath,
  amssymb,
  longbibliography,
  floatfix
]{revtex4-2}

\usepackage{graphicx}   
\usepackage{dcolumn}    
\usepackage{bm}         
\usepackage{array}      
\usepackage{tabularx} 
\usepackage{amsmath}
\newcolumntype{Y}{>{\centering\arraybackslash}X}
\usepackage[table]{xcolor}
\definecolor{lightblue}{RGB}{220,240,255}
\usepackage{xcolor}
\usepackage{soul}
\sethlcolor{yellow}
\usepackage{braket}     
\usepackage{comment}

\usepackage{hyperref}

\begin{document}


\title{Arbitrary control of the temporal waveform of photons during spontaneous emission}

\author{Carl Thomas}
\email{carl.jacob.thomas@gmail.com}
\affiliation{Department of Physics, University of Washington}
\author{Rebecca Munk}
\affiliation{Department of Physics, University of Washington}
\author{Boris B. Blinov}
\affiliation{Department of Physics, University of Washington}

\date{\today}

\begin{abstract}
Control of the temporal waveform of photons produced during spontaneous emission from single quantum emitters provides a crucial tool in the establishment of hybrid quantum systems, optimization of quantum state transfer protocols and mitigation of effects due interferometric instability for network architectures based on flying qubits. We describe a method to generate photons of any temporal waveform from emitters of any excited state lifetime, limited only by the timing resolution of control hardware. We show how the temporal waveform of photons can be controlled by deterministically varying the population of an excited state which undergoes spontaneous emission.  Our broadly applicable approach has only two requirements for a candidate quantum emitter: modulation of the (1) amplitude and (2) relative phase of a field coupling a ground state to the excited manifold. We detail how to identify optimal excitation pulses by employing variational algorithms to feed back on atomic populations. Additionally, we develop Quantum Monte Carlo based tools to determine photon-number statistics and establish techniques  to identify optimal excitation strengths and post-selection thresholds for photon generation protocols. We situate our work in the context of other prior research on bespoke single photon sources and networking including post-emission pulse shaping, temporal gating and cavity-based methods. In comparison, our free-space process has greater flexibility in producing any waveform, requires less infrastructure, and can be readily applied across a wide range of quantum emitters. We discuss the applications and limits of this technique, including how increasing photon emission probabilities affects achievable temporal-mode overlap fidelities between emitted and target photon waveforms.
\end{abstract}

\keywords{photon shaping, trapped ion, quantum network, hybrid system}

\maketitle

\section{\label{sec:introduction}Introduction}

The ability to control the state of single photons produced during spontaneous emission represents a key tool for building quantum networks. For example, the production of indistinguishable photons from distinct qubit implementations would allow the application of established entanglement generation protocols to realize hybrid quantum systems \cite{olmschenk_quantum_2010, lilieholm_photon-mediated_2020}. Such systems could leverage the strengths of various qubit types such as the long coherence times of ions or the fast gate speeds of solid state spin qubit to out-compete any single-qubit type platform \cite{chu_hybrid_2021}. To realize such a hybrid system, photons from each node must be identical in all bases including frequency, polarization, number, spatial mode and time bin \cite{beukers_remote-entanglement_2024}. As such, developing tools to engineer the photon temporal waveform proves crucial for enabling mixed architecture quantum systems. 

Same qubit-type quantum networks also benefit from tailoring photon temporal distributions to reduce errors and to maximize  entanglement generation rates. Additionally, characterizing photon emission statistics during an excitation process allows optimization of experimental parameters for network protocols. By controlling for the likelihood of multi-photon emission events during excitations, quantum network operators can reduce infidelities in remote entanglement experiments and use conditional photon detection statistics to perform improved post-selection. In this work we (i) present a method for arbitrary control of the temporal waveform of photons emitted by network nodes and (ii) develop modeling tools to quantify and reduce the impact of multiple excitations within a single trial.

Control over single-photon degrees of freedom such as frequency, polarization, and spatial mode is comparatively mature. Quantum frequency conversion provides a promising route for bridging spectral mismatch between disparate emitters \cite{siverns_neutral-atom_2019, yu_efficient_2025}. Polarization can be engineered by mapping the emitter's dipole radiation pattern onto well-defined laboratory-frame modes and using standard polarization optics (e.g., waveplates and polarizing beam splitters) \cite{blinov_observation_2004}. Spatial-mode control can be achieved using optical cavities \cite{mckeever_deterministic_2004, knall_efficient_2022}, fiber-based waveguides \cite{stephenson_high-rate_2020}, or aperture-based masking \cite{sosnova_mixed-species_nodate}; additionally, spatial light modulators offer a reconfigurable platform for tailoring mode profiles across a range of protocols \cite{faorlin_controlling_2025}.

In tandem with established control of frequency, polarization, and spatial mode, the ability to arbitrarily engineer the \emph{temporal} mode would remove a major barrier to interfacing heterogeneous quantum emitters. By matching the photon emission probabilities of both nodes, various single \cite{slodicka_atom-atom_2013, hermans_entangling_2023, beukers_remote-entanglement_2024, cabrillo_creation_1999} and two \cite{duan_scalable_2004} photon detection methods, well established in same-type qubit networks, could be used to entangle mixed-type qubit systems. Photon-based remote entanglement techniques have been independently implemented in neutral atoms \cite{van_leent_entangling_2022}, solid state systems \cite{bernien_heralded_2013} and trapped ions \cite{krutyanskiy_entanglement_2023}.  To our knowledge, comparable protocols have not yet been realized between \emph{distinct} emitter platforms. Interference between photons from a trapped ion and a (telecom-converted) neutral-atom ensemble has been demonstrated, but relied on temporal gating to obtain high-contrast interference \cite{craddock_quantum_2019}. If excitation pulse engineering (or other post-emission techniques) are not employed to ensure high indistinguishability of the temporal modes of photons from such mixed-architecture systems, the efficiency of gating will decrease as the difference between emitter lifetimes grows. This limitation motivates methods that directly engineer photon temporal modes to achieve high overlap without discarding most of the emitted light, reducing entanglement generation rates.

Full control of the temporal mode of single photons also enables optimization of quantum state-transfer processes in repeater architectures. For example, when the incident wavepacket is the time reverse of the spontaneous-emission wavepacket of a target transition, the absorption probability at a receiving node can be maximized \cite{stobinska_perfect_2009, golla_generation_2012}. Similarly, in atom--cavity systems, tailoring photon waveforms to impedance-match an optical resonator enhances coupling between flying and stationary qubits \cite{cirac_quantum_1997}.

Photon temporal waveform shaping can also be utilized to engineer wavepackets optimally robust to experimental error sources. For example, in trapped ion photonic interconnects, photons from distributed devices are most often produced via strong, short (relative to the emitter lifetime) excitation pulses \cite{saha_high-fidelity_2025, maunz_quantum_2007}. In such systems, the resultant photon waveform closely matches the Fourier transform--limited Lorentzian line-shapes of the excited state, yielding a temporal distribution with an exponentially decaying tail with the damping rate determined by the excited state lifetime. While photons from two such nodes will interfere perfectly assuming no interferometric error sources, the entangled state preparation fidelity for such experiments may be degraded by non-common mode path length fluctuations \cite{beukers_remote-entanglement_2024}. These deleterious effects can be minimized using photons with Gaussian waveforms which are the most robust to path length instability \cite{rohde_optimal_2005}. Analogously, timing jitter between excitation pulses applied to spatially separated nodes leads to similar effects, and its impact can likewise be mitigated by tailoring the photon temporal mode. As such, the ability to generate user-defined photon waveforms could prove useful in future long distance distributed quantum systems.

Challenges associated with hybrid networking, state transfer, and robustness to timing jitter and interferometric phase noise have motivated substantial prior work on controlling photon temporal waveforms. Existing approaches broadly fall into two categories: \emph{post-emission} methods that reshape a photon after it is generated, and \emph{in-situ} methods that engineer the waveform during the emission process. Post-emission tools include circuit based pulse shaping via dispersive optics \cite{sosnicki_interface_2023}, electro-optical modulation of photon wave packets \cite{kolchin_electro-optic_2008, wright_spectral_2017, sosnicki_large-scale_2018} and measurement based post-selection techniques \cite{lecamwasam_measurement-based_2017}. However, such approaches are not widely applicable across all frequencies owing to hardware limitations such as the general lack of dispersive elements in the UV band.  Moreover, these approaches can introduce additional insertion loss and technical noise. 

Approaches which aim to control photon waveform in-situ by embedding the emitter in a cavity have been widely implemented or proposed in many qubit platforms including ions \cite{keller_continuous_2004}, quantum dots \cite{you_developing_2025} and neutral atoms \cite{covey_quantum_2023}.  In cavity-assisted stimulated Raman adiabatic passage (STIRAP) schemes, shaped control fields drive a Raman process that emits a photon into a cavity mode with a programmable temporal profile. These techniques can generate a wide class of waveforms, but are constrained by adiabaticity requirements and by cavity parameters (e.g., linewidth and coupling), which set a minimum temporal feature size in the emitted wavepacket.  Such approaches also require specialized cavity infrastructure and efficient mode matching into the resonator \cite{vasilev_single_2010, kuhn_deterministic_2002, nisbet-jones_photonic_2013}, which can limit achievable photon generation rates and complicate scaling in already crowded experimental platforms. Moreover, while emission into a single cavity mode is advantageous for many networking tasks, it can be less compatible with protocols that encode atom--photon entanglement in multiple orthogonal modes (e.g., polarization or frequency-bin schemes) unless the resonator and level structure are engineered to support multiple emission pathways \cite{moehring_entanglement_2007}.

The technique we describe in this paper makes use of a free space quantum emitter wherein shaping of the photon temporal waveform is realized by deterministic modulation of the population of the state undergoing radiative decay. This control is realized by combining optimization of the pulse envelope (i.e. the amplitude of the time dependent Rabi frequency) with parity-phase control (i.e. advances of $\pi$ radians in the local oscillator phase) for a field which couples a ground and excited state. The conceptual approach, numerical methods to identify optimal excitation pulses and representative results for photon waveform shaping are discussed in Secs.~\ref{sec:any_emitter} and~\ref{sec:methods_results}. A brief comparison of the strengths, limitations and requirements of the various methods for photon temporal waveform control is provided in Appendix~\ref{app: temporal_control_comp}.

Beyond temporal-mode control, the performance of photon-mediated network protocols depends critically on single-photon purity at each node.  Finite-duration excitation pulses inevitably introduce a nonzero probability of multiple photon emissions from each node during a single entanglement attempt.  In practice, when the excitation pulse length is comparable to the excited-state lifetime, a photon emitted early in the pulse can be followed by re-excitation and a second emission within the same trial, even when the total emission probability is kept small.

These effects are generic across architectures based on spontaneous emission. Analyses of remote entanglement schemes explicitly show that multi-photon events set a trade-off between entanglement generation rate and fidelity: increasing the emission probability increases the heralding rate but also increases the likelihood of generating multiple photons from a single node during any given shot. Such events degrade the fidelity of the entangled state \cite{hofmann_heralded_2012, beukers_remote-entanglement_2024, hermans_qubit_2022}. It is therefore essential to (i) model the full photon-number statistics associated with a given temporal pulse shape, (ii) operate in regimes where the multi-photon probability is quantitatively controlled, and (iii) design temporal post-selection strategies that suppress contributions from trajectories involving more than one emission event. These considerations motivate the Quantum Monte Carlo (QMC) based tools developed in Sec.~\ref{sec:QMC_tools}, which we use to characterize and mitigate multi-photon error channels for shaped waveforms. We also describe how these Monte Carlo methods can be used to develop post-selection techniques which lead to improved entangled state fidelities. This is accomplished by using conditional temporal statistics to reject shots where detected photons are likely to have been generated during a multi-emission event.

\section{Quantum Optics Tools for Modeling the Temporal Mode of Photons}
\label{sec:any_emitter}

\begin{figure}[t]
  \centering
  \includegraphics[width=0.4\textwidth]{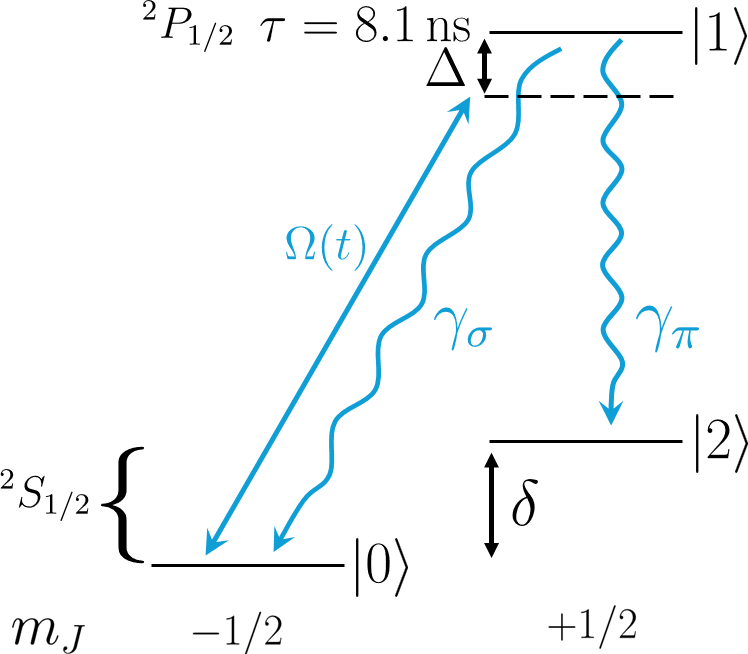}
  \caption{Partial level structure of $^{174}\text{Yb}^+$ relevant for photon control experiments. The application of a time-dependent Rabi frequency, $\Omega(t)$, drives the transition from $\ket{0}$ to $\ket{1}$ to control the excited-state population, $\rho_{11}(t)$. Spontaneous emission from $\ket{1}$ generates photons with a temporal waveform at rate $\Gamma\,\rho_{11}(t)$ in a superposition of $\sigma$ and $\pi$ polarizations with a probability ratio of 2:1.}
  \label{fig:levels}
\end{figure}

In this section, we establish the relevant quantum optical model to describe the generation of photons in a particular temporal mode during an atom-light interaction. As a concrete example, we consider the $^2\mathrm{S}_{1/2}$ and $^2\mathrm{P}_{1/2}$ manifolds of a $^{174}\mathrm{Yb}^+$ ion. By adapting the Hamiltonian describing the time evolution of the system, this approach can be applied to higher dimensional Hilbert spaces corresponding to more general multi-level qudits. We model an excitation process which drives the $(^2\text{S}_{1/2}, m_J=-1/2) \equiv \ket{0}\Rightarrow (^2\text{P}_{1/2}, m_J=+1/2) \equiv \ket{1}$ transition. State $\ket{1}$ decays back to $\ket{0}$, corresponding to the emission of a $\sigma$-polarized photon, and to $(^2\text{S}_{1/2}, m_J=+1/2) \equiv \ket{2}$, corresponding to the emission of a $\pi$-polarized photon. The relevant atomic levels are shown in Fig.~\ref{fig:levels}. 

The problem we aim to solve is to determine the appropriate form of the interaction Hamiltonian which captures the behavior of the field which couples states $\ket{0}$ and $\ket{1}$ such that the resultant population of the excited state, $\rho_{11}(t)$, follows the desired form for a particular target photon temporal distribution. The coupling field is parameterized by a time-dependent Rabi frequency $\Omega(t)$ and a time-dependent phase $\phi(t)$. Allowing $\phi(t)$ to vary in time will be important for enabling coherent de-excitation, as discussed below.

For a $\Lambda$-system driven by a laser pulse with an enveloped corresponding to a time-dependent Rabi frequency $\Omega(t)$, carrier frequency $\omega_L$ and controllable phase $\phi(t)$, after applying the rotating wave approximation, we can write the interaction Hamiltonian as
\begin{equation}
  H_{I}=\frac{\hbar}{2}\Omega(t)\Big(e^{i(\omega_L t+\phi(t))}\ket{0}\bra{1}+e^{-i(\omega_L t+\phi(t))}\ket{1}\bra{0}\Big).
\end{equation}
The atomic Hamiltonian simply captures the splitting between the excited state and ground states,
\begin{equation}
    H_{0}=\hbar\omega_0\ket{1}\bra{1},
\end{equation}
where $\hbar \omega_0$ is the energy of the excited state. We have initially assumed that the two Zeeman sublevels of the $^2\text{S}_{1/2}$ manifold are degenerate in energy as is the case when no external magnetic field is present. 
Thus, in matrix form, we have
\begin{equation}
    H_I+H_0= \frac{\hbar}{2} 
\begin{pmatrix}
0 & \Omega(t)e^{i(\omega_Lt+\phi(t))} & 0 \\
\Omega(t)e^{-i(\omega_Lt+\phi(t))} & 2 \omega_0& 0 \\
0 & 0 &  0
\end{pmatrix}.
\end{equation}

Adopting a frame which rotates at frequency $\omega_L$, we can easily include terms for time dependent laser detuning, given by $\Delta(t)=\omega_L(t)-\omega_0$ and any ground state splitting, $\delta$. This yields the rotating frame Hamiltonian
\begin{equation}
\tilde H(t) = \frac{\hbar}{2}
\begin{pmatrix}
0 & \Omega(t)\, e^{i\phi(t)} & 0\\[4pt]
\Omega(t)\, e^{-i\phi(t)} & -2\!\big[\Delta(t)+\dot{\phi}(t)\big] & 0\\[4pt]
0 & 0 & 2\delta
\end{pmatrix}.
\end{equation}

As we show in later sections, modulating the drive amplitude together with instantaneous $\pi$ phase flips of the coupling field is sufficient to generate photons of any temporal mode, limited only by the available control bandwidth. By a $\pi$ phase flip we mean that the drive phase (in the rotating frame) is restricted to $\phi(t)\in\{0,\pi\}$, such that the off-diagonal coupling terms in $\tilde H(t)$ change sign during the excitation pulse. Physically, a $\pi$ phase step reverses the sign of the coherent drive and therefore reverses the axis of the induced rotation between $\ket{0}$ and $\ket{1}$, enabling coherent de-excitation when desired. Experimentally, the required control of $\Omega(t)$ may be achieved by modulating the amplitude and phase of the driving field using standard acousto-optic (AOMs) or electro-optic modulators (EOMs). A more detailed discussion of experimental implementation, including consideration of modulator and power requirements, photon isolation and waveform validation, is provided in Appendix~\ref{app:exp_real}.

Having specified the coherent dynamics, we now include incoherent decay due to spontaneous emission from $\ket{1}$. For $^{174}\mathrm{Yb}^+$ we take the excited-state decay rate to be $\Gamma = 1/\tau \approx 1.23\times 10^{8}\,\mathrm{s}^{-1}$, with branching
\begin{equation}
\Gamma=\Gamma_{\sigma}+\Gamma_{\pi}, \qquad \Gamma_{\sigma}:\Gamma_{\pi}=2:1,
\end{equation}
so that $\Gamma_{\sigma}=2\Gamma/3$ and $\Gamma_{\pi}=\Gamma/3$. The corresponding collapse operators are
\begin{equation}
\hat L_{0}=\sqrt{\Gamma_{\sigma}}\,\ket{0}\!\bra{1}, \qquad
\hat L_{2}=\sqrt{\Gamma_{\pi}}\,\ket{2}\!\bra{1}.
\end{equation}
The density matrix then evolves according to the Lindblad master equation
\begin{equation}
\dot{\rho}=-\frac{i}{\hbar}\,[\tilde H(t),\rho]
+\sum_{k\in\{0,2\}}\left(\hat L_k \rho \hat L_k^{\dagger}
-\frac{1}{2}\left\{\hat L_k^{\dagger}\hat L_k,\rho\right\}\right)
\end{equation}
where $\rho$ is the density matrix for the atom. In matrix form, the dissipative contribution $\mathcal{L}(\rho)=\sum_k(\cdots)$ in the basis $\{\ket{0},\ket{1},\ket{2}\}$ is written as
\begin{equation}
\mathcal{L}(\rho)=
\begin{pmatrix}
\Gamma_{\sigma}\rho_{11} & -\frac{\Gamma}{2}\rho_{01} & 0\\[4pt]
-\frac{\Gamma}{2}\rho_{10} & -\Gamma\,\rho_{11} & -\frac{\Gamma}{2}\rho_{12}\\[4pt]
0 & -\frac{\Gamma}{2}\rho_{21} & \Gamma_{\pi}\rho_{11}
\end{pmatrix}.
\end{equation}

To simulate the time-evolution of $\rho_{11}(t)$ and thus determine the temporal mode of resultant photons, we employ the Python-based quantum optics library, QuTip, to numerically solve the Lindblad master equation with initial state  $\ket{0}$.

If we consider an excitation pulse of finite duration, we can describe the relative temporal flux, $I_q(t)$, of photons emitted into polarization mode $q\in\{\pi,\sigma\}$ as 
\begin{equation}
I_{q}(t)\ = \Gamma_{q}\,\rho_{11}(t).
\end{equation}
Normalizing $I_q(t)$ yields the temporal intensity envelope of the photons emitted during this process:
\begin{equation}
    |g_q(t)|^2=\frac{\Gamma_q\,\rho_{11}(t)}{\langle N_q \rangle},\qquad \int |g_q(t)|^2\,dt=1
\end{equation} 
where $g_q(t)$ is the temporal mode.
Here, $\langle N_q \rangle$ is the mean photon emission number into polarization mode q given by:
\begin{equation}
\label{eqn: emission_probability}
    \langle N_q \rangle\equiv \int_{t_0}^{t_f}\!\Gamma_q\,\rho_{11}(t) dt.
\end{equation}

For the case of single-photon emissions, we can write the state of the resultant photons using polarization-time mode creation operators,
\begin{equation}
\ket{1_{g_{q}}^{(q)}}\;=\;\int\! dt\; g_{q}(t)\; \hat a^\dagger_{q}(t)\ket{vac},
\label{eq:temporal_mode_def}
\end{equation}
where $\hat{a}_q^{\dagger}(t)$ is the creation operator for a photon in polarization mode, $q$, at time, $t$. 

Because the emission of a $\pi$ photon projects the atom into a dark state (i.e. the driving field does not couple $\ket{1}\leftrightarrow\ket{2}$), we have the condition $\langle N _\pi\rangle \le 1$. Therefore, the state of the $\pi$-polarized field is described by a two-dimensional subspace spanned by  $\{|\mathrm{vac}\rangle,\;|1^{(\pi)}_{g_\pi}\rangle\}$ with $P_0=1-\langle N_\pi\rangle$ and $P_1 = \langle N_\pi\rangle$ respectively. By contrast, a $\sigma$ emission returns the atom to $\ket{0}$, allowing re-excitation; the $\sigma$ field can therefore contain multiple photons occupying distinct temporal modes, and it is not generally restricted to a two-dimensional Fock subspace. 

To connect the atomic dynamics to a Fock-state description of the emitted photon, it is useful to pass from the atomic density matrix given by solving the master equation to the joint atom--field state. In the weak-excitation regime, where re-excitation following a $\sigma$ decay is negligible, the field Hilbert space may be truncated to the vacuum and single-photon bases. At late times $t$, after the excitation pulse has ended and any residual excited-state population has decayed, the joint atom--field state may then be written as
\begin{equation}
\begin{aligned}
\ket{\Psi(t)}
=\;&
\sqrt{1-\langle N\rangle}\,\ket{0}\ket{\mathrm{vac}}
+\sqrt{\langle N_\sigma\rangle}\,\ket{0}\ket{1^{(\sigma)}_{g_\sigma}}
\\
&+e^{-i(\delta t+\phi_0)}\sqrt{\langle N_\pi\rangle}\,\ket{2}\ket{1^{(\pi)}_{g_\pi}},
\end{aligned}
\label{eq:atom_photon_state_main}
\end{equation}
where $\langle N\rangle=\langle N_\sigma\rangle+\langle N_\pi\rangle\le 1$. The phase multiplying the $\ket{2}$ branch arises from the splitting $\delta$ between $\ket{0}$ and $\ket{2}$ in the $\{\ket{0},\ket{1}\}$ rotating frame, while $\phi_0$ encodes the initial phase of the driving field at $t=0$. An extended discussion of the atom--photon wavefunction at all times, including the implications of relative phase accumulation between $\ket{0}$ and $\ket{2}$, is given in Appendix~\ref{app:atom_photon_phase}.

In principle, the joint atom--field state can always be represented as a pure state in an expanded Hilbert space. However, once multi-photon emission is allowed---even in the restricted case where at most one $\pi$ photon can be produced while an arbitrary number of $\sigma$ photons may be emitted---the state is naturally expressed as a superposition over quantum-jump records (emission histories) with different photon numbers, times, and polarizations. For example, the atom may emit $n_{\sigma}=0,1,2,\ldots$ $\sigma$ photons at different times before a terminating $\pi$ emission (or no $\pi$ emission at all), and each trajectory leaves the atom in a different conditional state correlated with the emitted field. The exact joint state is therefore a highly entangled sum over all allowed $(n_{\sigma},n_{\pi})\in\{0,1,2,\ldots\}\times\{0,1\}$ and associated time records, and cannot be specified solely by the mean photon numbers $\langle N_{\sigma}\rangle$ and $\langle N_{\pi}\rangle$.

Identifying excitation pulses for which the `at-most-one-photon' truncation accurately represents the true joint atom--field state is nontrivial. The validity of this single-photon approximation depends not only on the mean emitted photon number $\langle N\rangle$ and intrinsic properties of the emitter (i.e. the excited state lifetime and branching ratio), but also the temporal profile of the driving field. In particular, $\langle N\rangle$ alone does not determine the probabilities of higher-order events $P(N\ge 2)$ nor the temporal correlations between emission times in such events. For example, even when $\langle N\rangle$ is small (e.g., $\langle N\rangle<0.2$), the probability of two or more emissions is nonzero and can vary substantially between excitation waveforms. In Sec.~\ref{sec:QMC_tools} we describe numerical  methods for computing photon-number statistics and time-resolved correlations for a given excitation pulse.

\section{Methods and results for photon shaping}
\label{sec:methods_results}

Having established the formalism for computing the temporal mode produced by a prescribed driving field, we now describe how we identify control fields that generate photons with a user-specified target waveform. Our pulse-design procedure is:
\begin{enumerate}
    \item \textbf{Initialization and amplitude scaling.} We choose a target temporal mode and a desired mean emission number. As an initial guess, we set the Rabi-frequency envelope to follow the target photon shape, $\Omega_0(t)\propto |g_{\mathrm{target}}(t)|$, and solve the master equation to obtain the resulting $\langle N\rangle$. We then scale the overall amplitude of $\Omega_0(t)$ until the computed emission number matches the target value, using a bracketed root-finding routine (Brent's method).
    
    \item \textbf{Objective function.} For the scaled initial pulse, we compute the overlap between the predicted and target temporal modes to define an optimization objective. Depending on the application, one may wish to optimize (i) the temporal-mode overlap alone, or (ii) a joint objective that incorporates both temporal-mode overlap and a target emission probability.
    
    \item \textbf{Variational optimization.} Starting from the initial guess, we iteratively update the time-dependent drive $\Omega(t)$ (and, when used, phase-parity control) to maximize the chosen objective. We employ either simulated annealing or covariance-matrix adaptation evolution strategy (CMA-ES) for this optimization.
\end{enumerate}

A central constraint is that, for a fixed target waveform, it is not generally possible to simultaneously prescribe an arbitrary temporal mode \emph{and} an arbitrary emission probability. When both the waveform and emission probability are specified, the achievable temporal-mode overlap depends on intrinsic emitter parameters and on the target distribution. We derive an upper bound on achievable values of $\langle N\rangle$ for a given target temporal mode $g(t)$ in Appendix~\ref{app:N_limits}. To contextualize the ability to generate photons across a range of distributions and the associated limits on waveform shaping, we reference four representative classes of target waveforms (`natural,' `long,' `short,' and `discontinuous') defined in Appendix~\ref{app:photon_classes}.

Initial waveform-optimization attempts relied solely on modulating the instantaneous Rabi-frequency amplitude, $\Omega(t)$, to match target photon temporal modes. We find numerically that amplitude modulation alone is sufficient to generate a broad class of \emph{long} temporal modes with high mode-overlap fidelity over a wide range of target emission probabilities. However, as the target wavepackets become progressively shorter relative to the excited-state lifetime, the amplitude-only approach breaks down. In particular, short or discontinuous target temporal modes cannot, in general, be reproduced with high fidelity using control of $\Omega(t)$ alone.

Without the ability to switch the direction of population transfer from $\ket{0} \rightarrow \ket{1}$ to $\ket{1} \rightarrow \ket{0}$ part way through a pulse, the rate at which the falling edge of a photon distribution decreases is limited by the excited state lifetime. The only exception to this photon distribution falling edge `speed-limit' is in the case that the integral of the excitation pulse Rabi frequency corresponds to a rotation on the Bloch sphere of the reduced $\{\ket{0},\ket{1}\}$ subspace greater than $R_x(\theta) = \pi$. In such a case, the falling edge of the photon distribution then receives an extra `kick' in decay rate as the driving field and contribution of spontaneous emission now pump population in the same direction.  However, even in these cases, the temporal distribution of the photon at later times still becomes dominated by the lifetime-limited exponential decay owing to non-negligible components of the photon temporal waveform arising from incoherent re-excitations and intrinsic dephasing due to spontaneous emission. In other words, in the case of resonant excitation wherein $\Delta =0$, amplitude-only modulation of the instantaneous Rabi frequency only permits state trajectories which trace out great circles on the Bloch sphere. A slightly expanded explanation of the reasons for these limitations, and additional information on temporal control limits based on intrinsic emitter properties, are described in Appendix~\ref{app:notes_on_temporal_control}.

Setting aside the experimental difficulty of achieving the large Rabi frequencies required to approximate short or discontinuous targets using amplitude-only control, this approach faces two limitations. First, as already noted, the late-time waveform typically retains a lifetime-limited exponential tail due to spontaneous-emission--induced dephasing. Second, attempting to enforce a rapid suppression of the photon flux over any portion of the waveform generally requires stronger overall driving, which increases the integrated excitation and reduces the ability to independently tune the mean photon emission probability to a low target value. Consequently, amplitude-only recipes for short wavepackets are poorly suited to protocols that require small emission probabilities, such as single-photon remote-entanglement schemes. Moreover, because re-excitation during the pulse cannot be quantitatively controlled in this setting, the probability of multi-photon events is no longer a `knob' available to a network user. This can lead to deleterious effects of higher order emission events which degrade performance in both single- and two-photon entanglement protocols.

To overcome the limitations of amplitude-only control, we introduce discrete phase control of the driving field within a single excitation pulse. Experimentally, this corresponds to imposing an (ideally) instantaneous phase step on the optical drive, for example via acousto or electro-optic modulation of the local-oscillator phase. The relevant hardware requirements for phase control are addressed in Appendix~\ref{app:exp_real}. In combination with amplitude modulation, restricting the drive phase to $\phi(t)\in\{0,\pi\}$ (a $\pi$ phase flip) allows the effective atom--light coupling to switch sign during the pulse, enabling coherent de-excitation and thereby access to short and discontinuous target temporal modes.

\begin{figure*}[ht]
    \centering
    \includegraphics[width=1\textwidth]{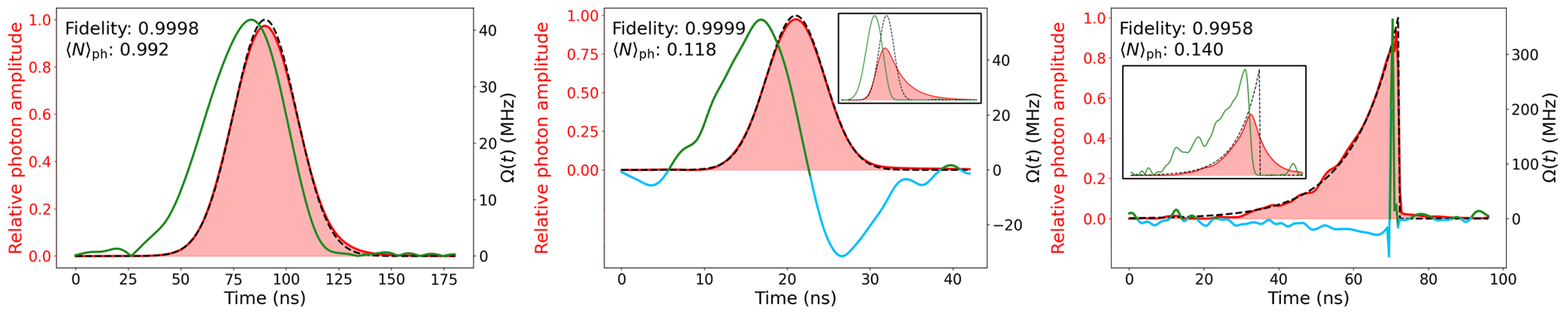}
    \caption{Representative numerically predicted photon waveforms and optimized pulses for a $^{174}\rm{Yb}^+\ (\tau =8.1\rm{ns})$. In all plots, green and blue solid lines represent in and out of phase (alternatively positive and negative) Rabi frequency amplitudes with the corresponding axis on the right hand side of each sub-figure. Dashed black lines show target photon distributions and filled red area show predicted photon shapes. \textbf{Left: } `Long'  Gaussian photon ($\sigma = 15$ns) demonstrating ability to produce waveforms with high fidelities and high ⟨N⟩ values without the use of phase modulation.  \textbf{Center: }   
    A `short' Gaussian photon ($\sigma = 3.5$ns). \textbf{Right: } An exponentially rising photon with $\tau_{\text{rise}} = 12 \text{ ns}$. For the center and right hand panes, insets show the best waveform which could be generated without the use of phase switching. Other representative photon waveforms, including flat-topped and triangular distributions can be found in Ref.~\cite{thomas_engineering_2025}.}  
    \label{fig:photon_shaping_results}
\end{figure*}

The effect of this phase-parity control is intuitive in the Bloch-sphere picture of the reduced $\{\ket{0},\ket{1}\}$ manifold. With a fixed phase $\phi=0$ and no time-dependent detuning, a time-dependent Rabi frequency drives the state along a great circle corresponding to an $R_x$ rotation. A $\pi$ phase flip is equivalent to reversing the sign of the drive (or, equivalently, applying a $\pi$ rotation about $\hat z$), which reverses the direction of the induced rotation on the Bloch sphere and allows population to be driven back from $\ket{1}$ to $\ket{0}$ on demand. In an ideal two-level system without dephasing, this provides a simple controllability argument: by choosing the instantaneous drive speed $\Omega(t)$ and its sign via $\phi\in\{0,\pi\}$, one can synthesize an arbitrary desired trajectory $\rho_{11}(t)$, and hence photons of any temporal modes.

In realistic emitters, spontaneous emission introduces decoherence that limits the achievable temporal-mode overlap fidelity, $\mathcal{F}$, for a given target waveform and mean emission number, $\langle N\rangle$. This interplay depends on both the emitter parameters and the target shape. While the relationship between the maximum $\mathcal{F}$ and for a given $\langle N\rangle$ cannot be analytically captured for an arbitrary target photon, we can qualitatively describe the factors which contribute to this interplay and provide some illustrative results for specific photon distributions in the context of the $\text{Yb}^+$ system we consider. 

Figure~\ref{fig:photon_shaping_results} shows representative target and optimized photon waveforms obtained using amplitude-only modulation and  combined amplitude modulation with $\pi$ phase flips. For long targets, amplitude-only control can achieve high temporal-mode overlap, whereas short and discontinuous targets require phase-parity control to enable coherent de-excitation. Appendix~\ref{app:hyrbid_pulses} further illustrates how emitters with different excited-state lifetimes can be driven to generate photons with the same temporal mode, enabling waveform matching for hybrid-network interfaces.

\begin{figure}[ht]
    \includegraphics[width=0.45\textwidth]{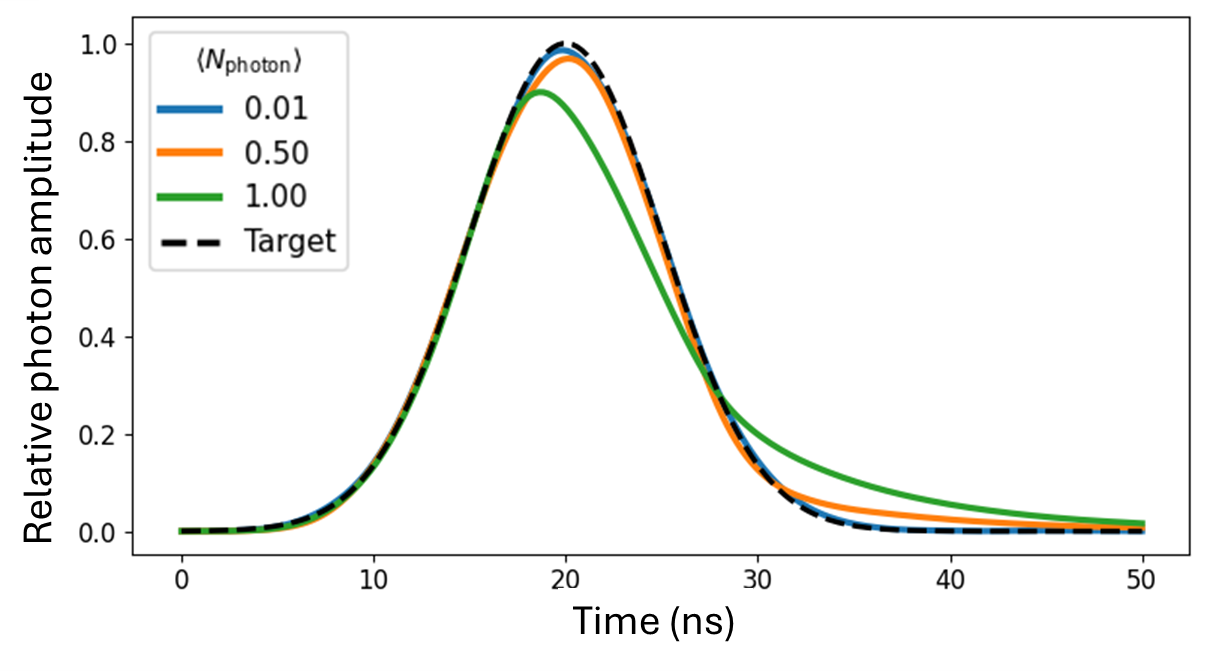}
    \caption{The best possible photon temporal waveforms which can be produced by $^{174}\rm{Yb}^+$ as target $\langle N \rangle$ is increased for a short Gaussian photon ($\sigma =5$ ns). Incoherent re-excitation leads to greater weight in photon tail at later times. These effects are suppressed in long photon temporal distributions.}
    \label{fig:vary_N}
\end{figure}

As stated, even with phase-parity control, incoherent re-excitation limits the maximum achievable temporal-mode preparation fidelity at a given mean emission number. For some given photon distribution, as $\langle N\rangle$ increases, a larger fraction of the emitted field originates from population that has already undergone an emission-and-return cycle during the pulse (i.e. via $\ket{1}\!\to\!\ket{0}$ decay), enabling subsequent re-excitation events. Re-excitation of this population is fundamentally an incoherent process owing to the effects of dephasing during initial emission. As such,  coherent de-excitation of some fraction of the excited state population is not possible, leading to a contribution to the photon waveform arising from the lifetime-limited exponential tail. Figure~\ref{fig:vary_N} illustrates this trade-off by showing the best achievable  waveform for a short target photon as $\langle N\rangle$ is varied. If $\langle N\rangle$ is left unconstrained, the optimizer can suppress re-excitation by selecting a low-emission solution, yielding arbitrarily high temporal-mode overlap at the cost of a reduced photon-generation probability. These effects can also be mitigated by choosing level schemes that reduce the branching probability for decay back to the driven ground state (i.e., suppressing the amplitude of $\ket{1}\!\to\!\ket{0}$ decay relative to $\ket{1}\!\to\!\ket{2}$).

\section{Quantum Monte Carlo tools for pulse optimization and post-selection}
\label{sec:QMC_tools}
Solving the Lindblad master equation as described in Sec.~\ref{sec:any_emitter} determines the temporal distribution of photons and mean photon emission number produced by a prescribed excitation pulse, but this alone does not yield the full photon-number statistics. In particular, neither $g(t)$ nor $\langle N\rangle$ determines the probability of multi-photon events or the time correlations between successive emissions. In multi-photon trials, emission times of subsequent photons are distinct but correlated, occupying different temporal modes. 

Quantitatively characterizing these statistics enables (i) selecting pulse strengths that optimize the photon generation rate versus single photon purity trade-off in remote-entanglement protocols and (ii) designing time-resolved post-selection rules that reduce systematic errors by rejecting photons likely to have been produced during multi-photon emission events. We outline how quantum-jump (quantum-trajectory) Monte Carlo simulations can be used to generate emissions statistics to address these challenges. In short, we simulate the evolution of a given excitation many times and treat spontaneous-emission as a discrete jump (rather than a continuous damping as in the prior sections). We record the time stamps and ordering of jumps (first photon, second photon, etc.) for each trajectory to reconstruct the relative frequencies and conditional temporal distributions of multi-photon emission events.

Table~\ref{tab:monte_carlo_photon_stats} summarizes photon-number statistics obtained from QMC simulations for several values of $\langle N\rangle$ for excitation pulses targeting a Gaussian temporal mode with $\sigma=15~\mathrm{ns}$. As an illustrative application of these statistics, consider a two-photon remote-entanglement protocol in which performance depends on generating a single photon from each node while suppressing multi-photon emission events that lead to false heralds and reduced fidelity. A useful single-node metric is the single-photon purity defined as the probability that a shot producing at least one detected photon contained exactly one emission from a given node. Denoting $P_n \equiv P(N=n)$, this purity is
\begin{equation}
\mathcal{P}_1 \equiv P(N=1\mid N\ge 1)=\frac{P_1}{\sum_{n\ge 1}P_n}=\frac{P_1}{1-P_0}.
\end{equation}

Increasing the drive strength generally increases $P_1$ but also increases the incidence of multi-photon events, thereby reducing the single-photon purity $\mathcal{P}_1$. In optimizing parameters for a network-protocol, $\mathcal{P}_1$ is treated as a protocol-imposed constraint tied to an error budget for multi-photon contamination. In the case of a two-photon remote entanglement scheme, a user can a target temporal mode $g(t)$ and use such a QMC simulation to determine the largest excitation strength (equivalently, the largest $\langle N\rangle$ within a given pulse family) that (i) achieves the desired temporal-mode overlap and (ii) satisfies a requirement such as $P_{\ge 2}\le \epsilon$, where  $\epsilon$ is the tolerated multi-photon contribution.

\begin{table}[tbp]
\centering
\begingroup
\footnotesize
\renewcommand{\arraystretch}{1.15}
\setlength{\tabcolsep}{4.5pt}

\begin{tabular*}{\linewidth}{@{\extracolsep{\fill}}ccccccc@{}}
\hline\hline
$\langle N\rangle$ & $P(0)$ & $P(1)$ & $P(2)$ & $P(\ge 3)$ & $\mathcal{P}_1$ & $t_c$\\
\hline
0.01 & 99.00\% & 1.00\% & 0.00\% & 0.00\% & 1.0000 & --\\
0.25 & 76.09\% & 23.12\% & 0.77\% & 0.01\% & 0.9671 & --\\
0.50 & 53.43\% & 43.22\% & 3.26\% & 0.08\% & 0.9282 & 67\\
0.75 & 33.53\% & 58.31\% & 7.83\% & 0.33\% & 0.8773 & 57\\
1.00 & 16.86\% & 67.62\% & 14.56\% & 0.96\% & 0.8133 & 55\\
\hline\hline
\end{tabular*}

\caption{Photon-number statistics from QMC simulations for excitation pulses targeting a Gaussian temporal mode with $\sigma=15~\mathrm{ns}$, shown as a function of the mean emitted photon number $\langle N\rangle$. Each row is estimated from $1.2\times 10^{5}$ trajectories; for each value of $\langle N \rangle$, the photon distribution is centered at $60~\mathrm{ns}$. Columns $P(0)$, $P(1)$, $P(2)$, and $P(\ge 3)$ give the probability that a single excitation attempt produces 0, 1, 2, or at least 3 photons, respectively. $\mathcal{P}_1$ is included to illustrate how increasing $\langle N \rangle$ decreases single-photon purity. The final column reports the post-selection threshold time $t_c$ (ns), defined as the earliest time bin for which the ratio of later-emission counts to first-emission counts exceeds $0.1$. As expected, the reported mean photon number satisfies $\langle N\rangle=\sum_{n=0}^{\infty} n\,P(N=n)$ for each row, within the statistical uncertainty associated with the finite Monte Carlo ensemble.}
\label{tab:monte_carlo_photon_stats}
\endgroup
\end{table}

Beyond optimizing excitation pulse strengths, the same QMC tools enable improved time-based post-selection by estimating emission-time distributions conditioned on multi-photon events. The goal is to reduce the probability that second, third or later photons are mistakenly used to herald entanglement. As a concrete example, consider a single-photon remote-entanglement protocol that relies on detecting a $\pi$-polarized photon. With non-unit detection efficiency, trajectories in which an undetected $\sigma$ photon is emitted before the detected $\pi$ photon (i.e., a $\sigma$ decay followed by a $\pi$ decay) can still produce a single detector click but correspond to an incorrect heralding event for the stationary qubits. Using quantum trajectories, we reconstruct the ordered emission-time distributions for the first and subsequent photons and thereby estimate the conditional probability
\begin{equation}
\begin{split}
p_{\mathrm{multi}}(t) &\equiv
P\!\left(\text{N} \ge 2\,\middle|\,\right.\\
&\left.\pi\text{ photon detected at time }t\right).
\end{split}
\end{equation}

\begin{figure}[!ht]
    \centering
    \includegraphics[width=0.45\textwidth]{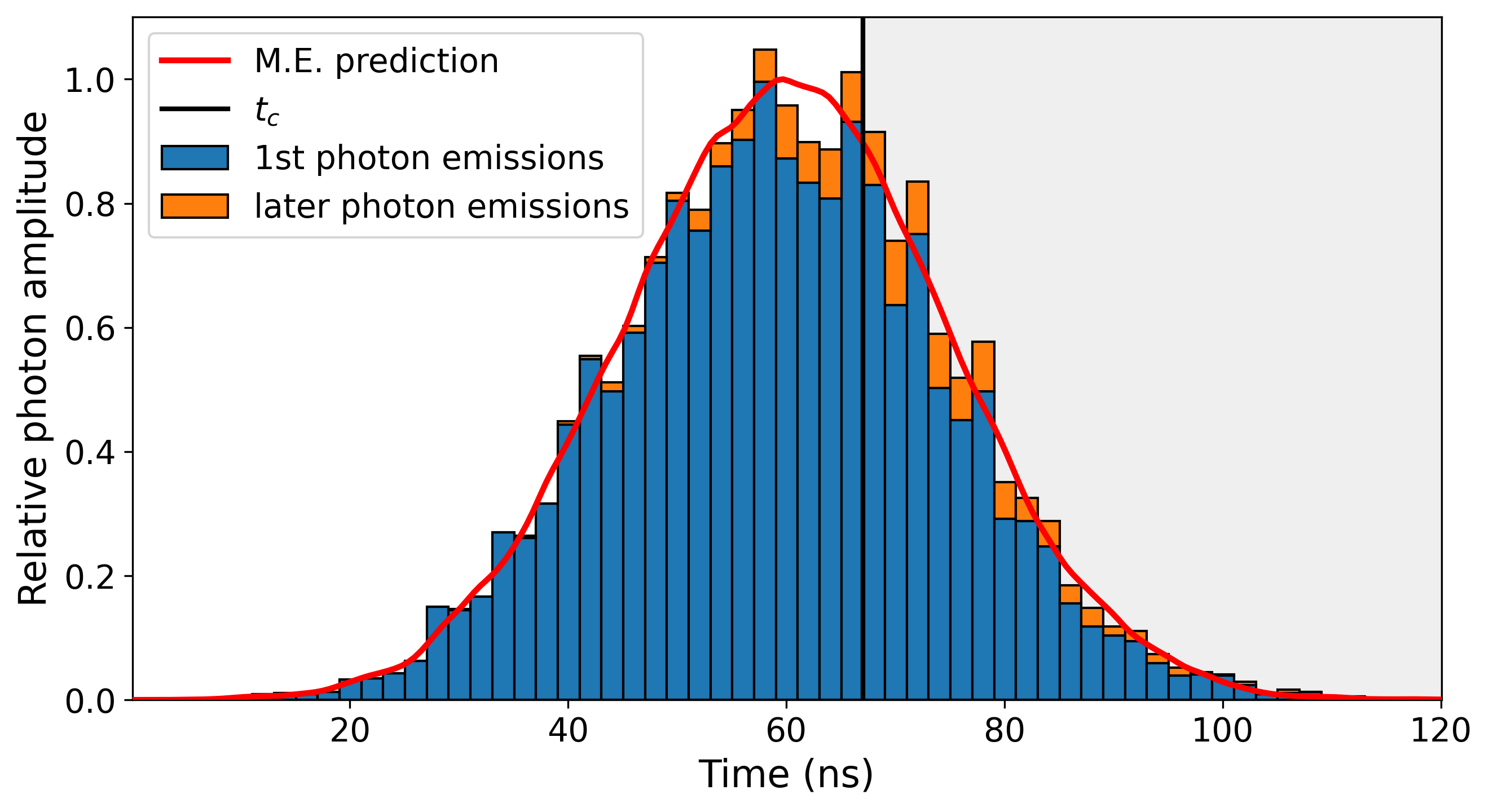}
    \caption{Photon emission-time distributions from Quantum Monte Carlo (QMC) simulations ($1.2\times 10^{5}$ trajectories) for an excitation pulse optimized to target a Gaussian temporal mode with $\sigma=15~\mathrm{ns}$ and mean emitted photon number $\langle N\rangle=0.5$. Blue bars show the emission-time histogram of the \emph{first} emitted photon in each trajectory, while orange bars indicate \emph{later} emissions (second and subsequent) in the same trial. The red curve is the corresponding master-equation prediction for the temporal distribution of emitted photons. The vertical line marks a threshold time $t_c$ defined as the earliest time bin for which the later-emission contribution exceeds $10\%$ of the first-emission contribution; such a time gate illustrates one post-selection strategy to suppress multi-photon contamination. Multi-photon statistics for these operating points are reported in Table~\ref{tab:monte_carlo_photon_stats}.}
    \label{fig:mc_histo}
\end{figure}

This enables a simple time-gating rule: choose a threshold $p_{\mathrm{th}}$ and discard detection events occurring after the cutoff time $t_c$ defined by $p_{\mathrm{multi}}(t_c)=p_{\mathrm{th}}$. Such post-selection improves the single-photon purity at the cost of reducing the heralding rate. Intuitively, for the $^{174}\mathrm{Yb}^+$ $\Lambda$-system driven on $\ket{0}\!\to\!\ket{1}$, a $\pi$ photon detected very early in the excitation window is unlikely to have been preceded by an undetected $\sigma$ emission, whereas late-time detections have a higher contamination probability. Figure~\ref{fig:mc_histo} illustrates this for a Gaussian target photon ($\sigma =15$ ns) at fixed $\langle N\rangle = 0.5$. The temporal distribution of first and all combined later photons are plotted and the threshold time corresponding to $p_{th}=0.1$ is displayed.  Table~\ref{tab:monte_carlo_photon_stats} reports the resulting cutoff times $t_c$ as $\langle N\rangle$ is varied for the same value of $p_{th}$.

A practical subtlety is that discarding detections after a cutoff time $t_c$ implements a trailing-edge time gate and therefore \emph{changes} the temporal distribution of the accepted events: the post-selected waveform is the conditional (renormalized) arrival-time distribution obtained by averaging only those trajectories that satisfy the time-of-arrival criterion. Consequently, post-selection must be designed so that the conditioned temporal modes from the two network nodes remain matched. For homogeneous nodes this is trivially achieved by applying the same cutoff rule to both arms. For hybrid nodes with different intrinsic lifetimes or drive constraints, matching will generally require distinct cutoffs chosen such that the \emph{post-selected} temporal distributions remain identical.

\section{Conclusion}
We have presented a free-space approach for engineering single-photon temporal modes by deterministically controlling the population dynamics of a radiative excited state during spontaneous emission. The central experimental requirements are the ability to modulate the amplitude of a driving field and to implement controlled phase steps (equivalently, sign changes of the atom--light coupling) on the $\ket{0}\!\leftrightarrow\!\ket{1}$ transition. Within the constraints set by finite control bandwidth, available Rabi frequency, and spontaneous-emission--induced decoherence and re-excitation, this method enables high-overlap synthesis of a broad class of target waveforms and can be applied to emitters spanning a wide range of lifetimes. These capabilities are directly relevant to interfacing heterogeneous quantum nodes, improving photon-mediated state transfer, and shaping wavepackets to mitigate timing and interferometric phase noise in remote-entanglement experiments. Finally, we showed how quantum-jump Monte Carlo simulations provide a practical framework for optimizing pulse strengths and designing time-based post-selection rules that suppress multi-emission error channels in photon-generation protocols.

\section{Acknowledgments}
This research was supported by the grant from the U.S.\ Department of Energy, Office of Science, Office of Basic Energy Sciences under Award No.\ DE-SC0020378. C.T.\ acknowledges helpful discussions with  A.~Hoyt and K.~M.~Fu.

\appendix 

\section{Comparison of methods for photon temporal waveform control}
\label{app: temporal_control_comp}

We summarize strengths, limitations, and experimental requirements of several approaches to single-photon temporal waveform control in Table~\ref{tab:photon_shaping}. No single method is optimal for all use cases. Cavity-based techniques are often preferred when near-deterministic single-photon generation into a well-defined spatial mode is the primary requirement. Temporal gating and other post-emission approaches are comparatively straightforward to implement, and can be effective when the emitter coherence time is short relative to the temporal resolution achievable with available modulators. Our approach is particularly well suited to photon-mediated networking protocols that benefit from reconfigurability, minimal additional infrastructure, and compatibility with a wide range of emitter lifetimes, subject to limits set by decoherence, re-excitation, and control bandwidth. Additionally, any system which meets the requirements of possessing a state which undergoes spontaneous emission which can be deterministically coupled to a stable or meta-stable manifold represents a candidate for photon temporal mode shaping in the manner we described. Indeed, the ability to produce a photon of a particular distribution will depend only on the available modulation speed, not on the emitter itself (albeit with different excitation pulse shapes). These considerations are addressed in Appendix~\ref{app:exp_real}. 

\begin{table*}[htbp] 
\caption{\label{tab:photon_shaping} Comparison of photon waveform temporal control methods.}
\centering
\label{tab:photon_shaping}
\begin{tabular}{llll}
\hline\hline
\textbf{Technique} &
\textbf{Strengths} &
\textbf{Limitations} &
\textbf{Requirements} \\
\hline
\parbox[t]{0.15\textwidth}{\raggedright Cavity STIRAP} &
\parbox[t]{0.26\textwidth}{\raggedright
Deterministic emission into a well-defined cavity mode; photon shaping via adiabatic control.} &
\parbox[t]{0.26\textwidth}{\raggedright
Requires high cooperativity; temporal features limited by adiabaticity and cavity dynamics; cavity coupling constraints.} &
\parbox[t]{0.26\textwidth}{\raggedright
High-finesse cavity; stable lasers; active cavity locking and mode matching. } \\[6pt]

\\ \parbox[t]{0.15\textwidth}{\raggedright Reshaping via time lenses} &
\parbox[t]{0.26\textwidth}{\raggedright
Flexible spectral/temporal reshaping with potentially high efficiency.} &
\parbox[t]{0.26\textwidth}{\raggedright
Phase-matching and bandwidth limits; dispersion loss; typically wavelength-specific hardware.} &
\parbox[t]{0.26\textwidth}{\raggedright
Dispersive media, high-speed phase modulators, synchronized waveform sources.} \\[6pt]

\\ \parbox[t]{0.15\textwidth}{\raggedright Temporal gating} &
\parbox[t]{0.26\textwidth}{\raggedright
Broadly applicable to many emitters; conceptually simple.} &
\parbox[t]{0.26\textwidth}{\raggedright
Discarding events reduces efficiency; performance degrades rapidly for large lifetime mismatch. } &
\parbox[t]{0.26\textwidth}{\raggedright
Fast detection and timing electronics; precise synchronization. 
} \\[6pt]

\\ \parbox[t]{0.15\textwidth}{\raggedright Post-emission modulation}  &
\parbox[t]{0.26\textwidth}{\raggedright
Reconfigurable waveform shaping; arbitrary shaping capability subject to hardware limits} &
\parbox[t]{0.26\textwidth}{\raggedright
Insertion loss; finite modulation depth; temporal resolution limited by modulator bandwidth; may add technical noise.} &
\parbox[t]{0.26\textwidth}{\raggedright
High-bandwidth modulators and drivers; stable RF/optical synchronization.} \\[6pt]

\\ \parbox[t]{0.15\textwidth}{\raggedright Direct radiative-state control (this work)}  &
\parbox[t]{0.26\textwidth}{\raggedright
Reconfigurable temporal-mode synthesis at the emitter; minimal added optical infrastructure. } &
\parbox[t]{0.26\textwidth}{\raggedright
Achievable fidelities and rates limited by spontaneous-emission decoherence, re-excitation, and finite control bandwidth/drive strength.} &
\parbox[t]{0.26\textwidth}{\raggedright
Control of the Rabi-frequency envelope and discrete phase steps (phase-parity control) of the excitation field.} \\ \\
\hline\hline
\end{tabular}
\end{table*}

\section{Notes on experimental realization: modulator and power requirements, photon isolation and waveform validation}
\label{app:exp_real}

In our approach, the time-dependent Rabi frequency is controlled by modulating the amplitude and phase of the driving field, with $\Omega(t)\propto E(t)$, where $E(t)$ is the electric-field amplitude of the excitation laser. Experimentally, this may be implemented using standard AOM or EOM techniques driven by programmable RF waveforms. The parity-phase protocol requires discrete $\pi$ phase shifts of the drive, which can be realized by advancing the phase of the RF source waveform; for both AOM- and EOM-based modulation, this phase update maps directly onto the optical phase of the excitation pulse and can be performed on timescales set by the inverse bandwidth of the device. As a result, any waveform generator/optical modulator combination capable of producing the desired intensity modulation will also have sufficiently fast phase control to reverse the direction of pumping along the $\ket{0}\leftrightarrow\ket{1}$ transition. The required control bandwidth is set by the characteristic timescale of the target waveform and limited by available hardware. As a practical rule of thumb, high-fidelity shaping typically requires a control resolution at least an order of magnitude shorter than the shortest relevant waveform feature. Commercial waveform generators and EOMs with bandwidths of tens of GHz are available, enabling control on sub-nanosecond timescales, while AOMs provide sufficient bandwidth for many of the waveform classes considered here.

In order to develop a qualitative sense of the control power needed to realize the waveforms presented in this text, we can connect the simulated Rabi frequencies to realistic experimental parameter. For the ${}^{174}\mathrm{Yb}^{+}$ ${}^{2}S_{1/2}\rightarrow{}^{2}P_{1/2}$ transition at $369~\mathrm{nm}$ used in our simulations, the dipole matrix element is of order $d \approx 1.5 \times 10^{-29}\,\mathrm{C\,m}$. Using this value, a Rabi frequency of $2\pi \times 100~\mathrm{MHz}$ corresponds to an electric-field amplitude of $\approx 4.5 \times 10^{3}\,\mathrm{V/m}$. For a Gaussian beam with waist $w_0 = 30~\mu\mathrm{m}$, this corresponds to an optical power of $\approx 37~\mu\mathrm{W}$ at the beam center; even $2\pi \times 500~\mathrm{MHz}$ requires less than $1~\mathrm{mW}$. These power levels are readily achievable with standard laser systems (such as extended cavity diode laser designs) and moderate focusing optics.

Experimental verification of the generated photon states may be performed using standard quantum optical techniques. The single-photon nature of the emission can be confirmed via measurements of the second-order correlation function $g^{(2)}(0)$ using a Hanbury Brown--Twiss setup. The temporal mode can be characterized using time-resolved photon counting to reconstruct $|g(t)|^2$, while interference-based measurements, such as Hong--Ou--Mandel interference with a reference photon, provide a direct probe of temporal-mode overlap relevant for quantum networking applications. In practice, separation of the driving field from the emitted photon can be achieved using standard techniques routinely employed in atomic and ion-based photon generation experiments. Possibilities include a combination of spectral, polarization, and/or spatial filtering. In particular, the emitted photons can be distinguished from the driving field via their spatial mode. In many qubit implementations such as ions and neutral atoms, fluorescence is collected in a direction orthogonal to the excitation beam. These methods are well established and sufficient to isolate the emitted photon from the driving field in typical implementations.  A description of experimental excitation pulse generation, emitted photon collection and temporal mode reconstruction can be found in Ref.~\cite{thomas_engineering_2025}.

\section{Expanded treatment of the atom--photon state and implications for network protocols}
\label{app:atom_photon_phase}

To clarify the phase appearing in Eq.~(\ref{eq:atom_photon_state_main}), it is useful to distinguish between the conditional (quantum trajectory) and unconditional (master equation) descriptions of the emission process. In the weak-excitation regime, where at most one spontaneous-emission event occurs, the joint atom--field state may be written as
\begin{equation}
\begin{aligned}
\ket{\Psi(t)}=\;&c_0(t)\ket{0}\ket{\mathrm{vac}}+c_1(t)\ket{1}\ket{\mathrm{vac}}
\\
&+\int_0^t dt'\,
\Big[
\psi_\sigma(t;t')\,\ket{0}\hat a_\sigma^\dagger(t')\ket{\mathrm{vac}}
\\&+\psi_\pi(t;t')\,\ket{2}\hat a_\pi^\dagger(t')\ket{\mathrm{vac}}
\Big],
\end{aligned}
\label{eq:joint_state_appendix}
\end{equation}
where $\hat a_q^\dagger(t')$ creates a photon in polarization mode $q\in\{\sigma,\pi\}$ emitted at time $t'$. In the rotating frame,
\begin{equation}
\begin{aligned}
\psi_\pi(t;t')&=\sqrt{\Gamma_\pi}\,c_1(t')\,e^{-i\delta (t-t')}e^{i\phi_0},
\\
\psi_\sigma(t;t')&=\sqrt{\Gamma_\sigma}\,c_1(t')
\end{aligned}
\label{eq:psi_appendix}
\end{equation}
Equation~(\ref{eq:joint_state_appendix}) is therefore the appropriate pure-state description of the atom and field during the excitation/emission process, truncated to at most one emitted photon.

At a final time $t_f$, after any remaining excited-state population has decayed, Eq.~(\ref{eq:joint_state_appendix}) reduces to Eq.~(\ref{eq:atom_photon_state_main}). Equivalently, the phase multiplying the $\ket{2}$ branch in Eq.~(\ref{eq:atom_photon_state_main}) may be absorbed into the temporal mode of the emitted $\pi$ photon. 

In the quantum-trajectory picture, a detection event at time $t_{\mathrm{det}}$ selects a particular emission history and yields a conditional state with a well-defined phase. If the photonic measurement erases which-path information by projecting onto a superposition of polarization modes, the conditional atomic state takes the form
\begin{equation}
\ket{\psi_{\mathrm{cond}}(t)}
\propto
\alpha\,\sqrt{\Gamma_\sigma}\,\ket{0}
+
\beta\,e^{-i[\delta (t-t_{\mathrm{det}})+\phi_0]}\sqrt{\Gamma_\pi}\,\ket{2},
\label{eq:conditional_atomic_state_main}
\end{equation}
where $\alpha$ and $\beta$ are determined by the projection of the photonic polarizations onto the lab-frame axis. In this  picture, the detection event establishes the reference time from which the $\ket{0}$--$\ket{2}$ superposition precesses.

For the temporal-mode shaping problem considered in this work, these phase considerations do not affect the calculated photon temporal modes, which depend only on $\rho_{11}(t)$. However, the phase structure of the emitted atom--photon state becomes relevant in networking protocols. For standard two-photon protocols with identical emitters, a coincidence event in the Bell-state analyzer heralds an atomic Bell state of the form
\begin{equation}
\ket{\Psi^\pm}
=
\frac{1}{\sqrt{2}}
\left(
\ket{0}_A\ket{2}_B
\pm
e^{i\phi_{\mathrm{det}}}
\ket{2}_A\ket{0}_B
\right),
\end{equation}
where the sign and the fixed phase offset $\phi_{\mathrm{det}}$ are determined by the detector geometry and any static calibration phases of the analyzer. In this ideal same-species case, the click-time difference does not itself generate the relative Bell-state phase. By contrast, when the two nodes acquire different dynamical phases (such as is the case for unequal $\ket{0}-\ket{2}$ splitting between nodes), the heralded state carries a conditional phase that depends on the full detection record and must be tracked or compensated. In single-photon protocols of the Cabrillo type, one must additionally account for optical path phase, so interferometric phase stabilization or active compensation is required \cite{cabrillo_creation_1999}.

\section{Limits on $\langle N \rangle$ for a target photon temporal distribution}
\label{app:N_limits}

We derive an upper bound on the achievable mean photon emission number for a specified target temporal \emph{intensity} profile by combining (i) the relation between photon flux and excited-state population and (ii) the physical constraint $\rho_{11}(t)\le 1$. To avoid confusion with the normalized single-photon mode amplitude $g(t)$ used in the main text, we write the target (nonnegative) intensity envelope as $f(t)\ge 0$ over the design window $t\in[t_0,t_f]$. When a normalized single-photon temporal mode is specified, one may take $f(t)=|g(t)|^2$.

In our model the instantaneous photon flux into polarization mode $q$ is given by Eqn. 9.
If we aim to realize a target flux proportional to $f(t)$, we may write the corresponding excited-state population profile as
\begin{equation}
\rho_{11}(t)= s\, f(t),
\end{equation}
where the scale factor $s$ is bounded by population conservation,
\begin{equation}
\begin{aligned}
s\,f_{\max} &\le 1, \qquad
s_{\max} = \frac{1}{f_{\max}}, 
\qquad 
f_{\max} \equiv \max_{t\in[t_0,t_f]} f(t).
\end{aligned}
\end{equation}

A conservative bound on the mean number of photons emitted \emph{within the finite window} $[t_0,t_f]$ follows by noting that population present in $\ket{1}$ at time $t$ decays before $t_f$ with probability $1-e^{-\Gamma(t_f-t)}$ (in the absence of further coherent control). This yields
\begin{equation}
\label{eq:Nq_window_bound}
\begin{aligned}
\langle N_q\rangle_{\max}\ \le\ 
\Gamma_q\, s_{\max}\!\int_{t_0}^{t_f}\! f(t)\,\Big[1-e^{-\Gamma(t_f-t)}\Big]\,dt
\ =\ \\
\frac{\Gamma_q}{f_{\max}}\int_{t_0}^{t_f}\! f(t)\,\Big[1-e^{-\Gamma(t_f-t)}\Big]\,dt.
\end{aligned}
\end{equation}
The bracketed term accounts for the fact that only a fraction of the excited-state population at time $t$ contributes to photons emitted before the end of the pulse window at $t_f$; any population that survives to later times would generate an additional lifetime-limited tail.

Equation~\eqref{eq:Nq_window_bound} is an upper bound based solely on population constraints and exponential decay, and it can be further tightened in practice by finite control bandwidth, bounded drive strength, and decoherence that limits how accurately $\rho_{11}(t)$ can be sculpted. Targets with larger early-time weight and emitters with longer optical coherence times (larger $T_2$) can approach the bound more closely, whereas strongly time-localized or discontinuous targets generally admit smaller achievable $\langle N_q\rangle$ for a fixed waveform fidelity.

\section{Description of photon classes}
\label{app:photon_classes}
We can categorize photon temporal distributions with the following classes:
\begin{enumerate}
    \item \textbf{Natural photons}: this is the photon generated by an excitation pulse of infinitely short duration. This yields a temporal distribution following an exponential decay set by the excited state lifetime. These are referred to as `natural' as this is the photon shape determined purely by the excited state line shape.
    \item \textbf{Long photons}: these are smooth monotonic photons with temporal spreads larger than natural photons.  
    \item \textbf{Short photons}: these are smooth monotonic photons with temporal spreads smaller than natural photons.  
    \item \textbf{Discontinuous photons}: these are photons whose temporal distributions include discontinuities, excluding the case of `natural' photons.     
\end{enumerate}

\begin{figure}[htbp]
  \centering
  \includegraphics[width=0.4\textwidth]{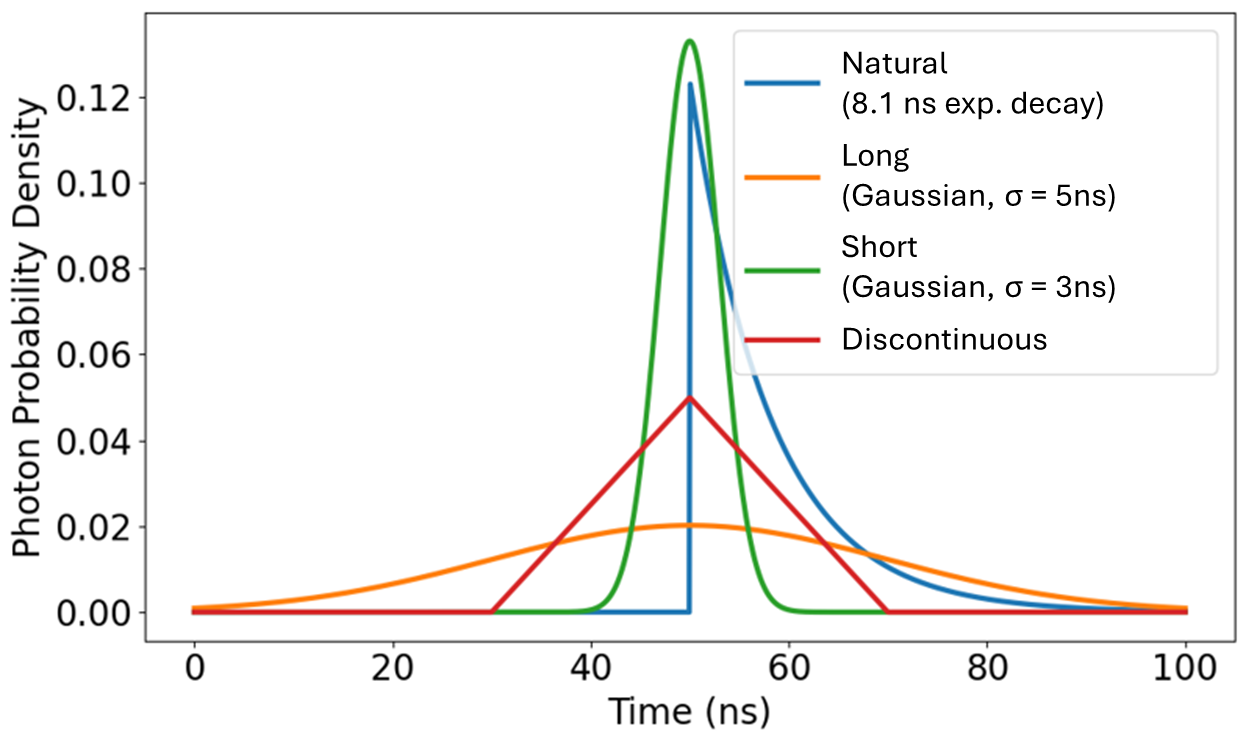}
  \caption{Representative, normalized temporal distributions for each photon class.}
  \label{fig: photon_classes}
\end{figure}

A representative photon temporal distribution for each class is shown in Fig.~\ref{fig: photon_classes}

\section{Notes on temporal control by photon class}
\label{app:notes_on_temporal_control}

The reduced performance of amplitude-only waveform synthesis for short and discontinuous targets arises from three related mechanisms:
\begin{enumerate}
\item \textbf{Re-excitation via cycling back to the driven ground state.}
Spontaneous emission that returns population to $\ket{0}$ during the pulse (e.g., $\ket{1}\!\to\!\ket{0}$) enables subsequent re-excitation and additional emission events. This contribution grows with the mean emission number $\langle N\rangle$ and is suppressed by level schemes that preferentially shelve population away from $\ket{0}$ (i.e., a small $\ket{0}:\ket{2}$ branching ratio), as well as by operating at lower $\langle N\rangle$.

\item \textbf{Optical decoherence from spontaneous emission.}
Even in the absence of technical noise, radiative decay induces dephasing of the $\{\ket{0}\!, \ \!\ket{1}\}$ coherence at a rate set by the excited-state relaxation, limiting how accurately the drive can deterministically sculpt $\rho_{11}(t)$ over extended durations. This loss of coherence effectively introduces incoherent population dynamics whose impact increases as $\langle N\rangle$ grows.

\item \textbf{Coherent drive direction and practical constraints.}
With a fixed drive phase (amplitude-only control) and no time dependent detuning, coherent evolution in the $\{\ket{0},\ket{1}\}$ manifold is constrained to rotations about a fixed Bloch-sphere axis, which prevents reversing the direction of population transfer mid-pulse. Producing sharp turnarounds, short fall times, or discontinuities in $\rho_{11}(t)$ therefore requires additional control degrees of freedom (e.g., discrete $\pi$ phase flips that change the sign of the coupling) and sufficient temporal bandwidth.
\end{enumerate}

For a given emitter, items (1) and (2) are fundamentally tied to spontaneous emission on the transition used for photon generation and therefore cannot be eliminated entirely; their quantitative impact depends on the chosen level structure, the target waveform, and the operating $\langle N\rangle$. In practice, these error channels can be reduced by working in a low-$\langle N\rangle$ regime when a lower photon-generation rate is acceptable, and by selecting transitions that minimize the probability of returning to the driven ground state.

\section{Generating photons for hybrid networks}
\label{app:hyrbid_pulses}
\begin{figure}[htbp]
    \includegraphics[width=0.45\textwidth]{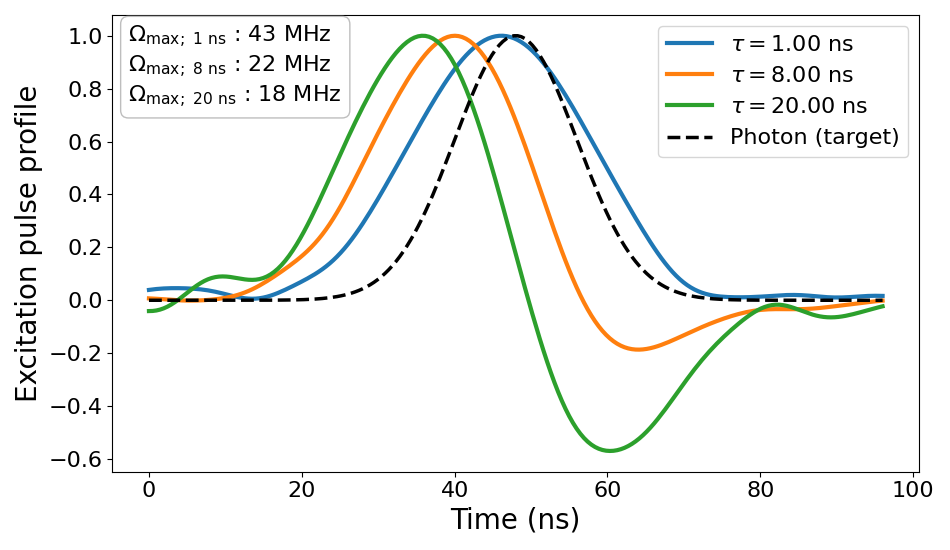}
    \caption{Optimized excitation pulse shapes to produce a desired Gaussian photon ($\sigma = 8 \text{ ns}$) for emitters of a variety of lifetimes. As $\tau / \sigma$ grows, the corresponding pulse shape shows earlier phase flips and greater contribution of out of phase contributions to $\Omega(t)$. For each emitter, the associated photon corresponds to an $\langle N \rangle$ of 0.10 and mode overlap between the predicted and target photon of at least $0.9995$.}
    \label{fig:multiple_emitters}
\end{figure}
Figure~\ref{fig:multiple_emitters} shows how identical photons can be generated from emitters of a range of lifetimes. Such photons could then be used to remotely entangle distinct qubits.


\input{article.bbl}

\end{document}

%% file: article.bbl
%